\documentclass[12pt]{iopart}

\usepackage{iopams}
\usepackage{graphicx}
\usepackage{pstricks}

\newcommand{\mod}{{\rm{mod}}}
\newcommand{\ket}[1]{|#1\rangle}

\newcommand{\bra}[1]{\langle#1|}

\newcommand{\of}[1]{(#1)}
\newcommand{\off}[1]{\left(#1\right)}
\renewcommand{\vec}[1]{\mathbf{#1}}

\begin{document}

\title[Quasi-angular momentum in rotating optical lattices]{Quasi-angular momentum of Bose and 
Fermi gases in rotating optical lattices}
\author{Brandon M. Peden, Rajiv Bhat, Meret Kr\"{a}mer, and Murray J. Holland}
\address{JILA, NIST and Department of Physics, University of Colorado, Boulder CO
80309-0440, USA}
\ead{pedenb@colorado.edu}

\begin{abstract}
The notion of quasi-angular momentum is introduced to
label the eigenstates of a Hamiltonian with a discrete rotational
symmetry. This concept is recast in an operatorial form where the
creation and annihilation operators of a Hubbard Hamiltonian carry
units of quasi-angular momentum. Using this formalism, the ground states 
of ultracold gases of non-interacting fermions in rotating optical lattices 
are studied as a function of rotation, and transitions between states of different 
quasi-angular momentum are identified. In addition, previous results for 
strongly-interacting bosons are re-examined and compared to the results for 
non-interacting fermions. Quasi-angular momentum can be used to distinguish 
between these two cases. Finally, an experimentally 
accessible signature of quasi-angular momentum is identified
in the momentum distributions of single-particle eigenstates.
\end{abstract}

\maketitle

\section{Introduction}

Recent experimental and theoretical studies of quantum gases of ultracold 
atoms have been successful in replicating the behavior of a wide class of 
condensed matter systems. In particular, Bose-Einstein condensation (BEC) 
\cite{anderson1995obe,davis1995bec,bradley1995ebe} 
and the degenerate Fermi gas \cite{demarco1999ofd} have both been observed, and there have been 
extensive studies of superfluidity, vortex formation \cite{zwierlein2005vas}, 
and the BCS-BEC crossover \cite{greiner2003emb,milstein2002rtc}. 
The Mott-insulator to superfluid phase transition in the Bose-Hubbard model has been 
observed in a system of ultracold bosons in an optical lattice of \cite{fisher1989sit,greiner2002qpt}. 
In addition, experiments observing novel phases of ultracold matter in atomic mixtures of 
Bose and Fermi gases in optical lattices have been performed 
\cite{albus2003mbf,illuminati2004hta,ospelkaus2006lba,ospelkaus2006uhm}.

Along these lines, focus has turned to strongly-correlated effects in systems found outside 
the realm of condensed matter. 
One promising line of research in the realization of these systems 
is that of coupled arrays of optical cavities, which allow single-site addressing 
\cite{angelakis2007pbi,hartmann2006sip,greentree2006qpt}.
Another is a mapping between the Hamiltonian of electrons confined in two 
dimensions in the presence of a constant transverse magnetic field and that of a 
rotating atomic gas. This has led to various theoretical studies predicting fractional quantum Hall 
effect (FQHE) behavior in rotating BEC's
\cite{wilkin2000ccb,paredes2001asa,fischer2004vlv,baranov2005fqh}. However, experimentally 
reaching the parameter regimes necessary to observe this behavior is difficult 
\cite{schweikhard2004rrb}. Optical lattices offer a solution to this problem by enhancing 
correlations. Optical lattices are formed as standing waves of 
counter-propagating laser beams and act as a lattice for ultracold atoms. These 
systems are highly tunable: lattice spacing and depth can be varied by tuning the 
frequency and intensity of the lasers, and interactions between atoms can be tuned 
via a Feshbach resonance \cite{bloch2005uqg}. 
Some recent 
studies have made direct connections between FQHE physics and strongly-interacting 
bosons in optical lattices in the presence of an effective magnetic field 
\cite{jaksch2003cem,palmer2006hff,sorensen2005fqh,bhat2007he,hafezi2007fqh}.

Connecting these theoretical studies to experiment requires identifying observables that 
can act as experimental signatures for the predicted physics. For instance, linear response 
theory has been applied to current flow in the presence of a ``potential gradient'' in 
order to observe FQHE physics \cite{bhat2007he}, noise correlation analysis has been 
applied to bosons 
in a rotating ring lattice \cite{rey2006llt}, and Bragg spectroscopy has been mentioned as a 
probe for vortex states \cite{vignolo2007qvo}. A common observable in ultracold atomic 
gases is the momentum distribution, measured in time-of-flight experiments.

In this paper, we present analytical results for strongly-interacting bosons in rotating 
optical lattices that confirm numerical results in a previous paper \cite{bhat2006bec} 
and extend them to systems of non-interacting fermions. These results are based 
on the notion of quasi-angular momentum, which 
is a quantum number for systems with a discrete rotational symmetry. Quasi-angular momentum 
is analogous to quasi-momentum for periodic translationally invariant systems 
\cite{bozovic1984pbs} and has been previously used in the context of rotating ring lattices 
\cite{buonsante2005aub,ueda1999gsp,buonsante2005aas} and carbon nanotubes 
\cite{reich2004cnb} to label eigenstates. A formalism for this quantum number for 
second-quantized systems is presented here and applied to gases of bosons and fermions in 
rotating optical lattices. The results presented identify transitions between 
states of different symmetry for the ground state of a rotating system. 
A possible avenue for experimentally detecting these results via the momentum 
distribution of the ground state is also presented.

The quasi-angular momentum, $m$, of the ground state of an ultracold
quantum gas of strongly-interacting bosons or non-interacting fermions in a
rotating ring lattice of $N$ sites can be monitored as a function of
rotation speed. For the case of strongly-interacting bosons, $m$ cycles through the values 
$m=nl~\mod~N:l=0,1,2,\dots$, where $n$ is the number of particles. For the
case of fermions, $m$ cycles through the values $m=nl~\mod~%
N:l=0,1,2,\dots$ for odd numbers of particles and $m=n(l+1/2)
~\mod~N:l=0,1,2,\dots$ for even numbers of particles. A system of non-interacting 
fermions is thereby distinguishable from a system of hard-core bosons. 
Similar behavior obains for systems of fermions in two-dimensional square lattices. Signatures 
of this quantum number in experiment are observable via the momentum
distribution of the ground state. For single-particle systems in two-dimensional square 
lattice geometries, such signatures include the existence of a peak at zero momentum only 
for the $m=0$ state and peak-spacing differences between $m=2$ and $m=1,3$ state.

The paper is structured as follows. In section 2, we draw the analogy
between systems with a discrete translational symmetry and systems with a
discrete rotational symmetry, thereby generating a Bloch theory for the latter 
and introducing the notion of quasi-angular momentum. Specific 
investigations of quantum gases of strongly-interacting bosons and non-interacting 
fermions in ring and square lattice geometries are carried out in section 3. In section 4, 
signatures of the quasi-angular momentum in the momentum distribution of a state 
are identified. Section 5 summarizes the main results of the paper.

\section{Quasi-angular momentum}

In this section, we briefly review band theory and introduce the language of
Bloch functions and quasi-momentum (see for instance \cite{marder2000cmp,kittel1996ssp}). 
By analogy, the \textit{quasi-angular momentum} of an eigenstate of a system with a 
discrete rotational symmetry is defined.

\subsection{Bloch theory}

When a small periodic potential of lattice 
period $d$ is introduced to a free-particle system, the two wavefunctions 
$\psi _{1}\left( x\right)=e^{i\pi x/d}$ and $\psi _{2}\left( x\right) =e^{-i\pi x/d}$ 
at the Brillouin zone boundaries ($q=\pm\pi/d$) mix with each other. The 
potential acts as a perturbation that breaks the degeneracy between these 
two states, creating new eigenfunctions,
\begin{equation}
\psi _{\pm }\left( x\right) =\frac{1}{\sqrt{2}}\left( e^{i\pi x/d}\pm
e^{-i\pi x/d}\right).
\end{equation}
This process opens up a band-gap at the Brillouin zone boundaries. 
In a reduced zone scheme, this energy spectrum becomes 
the familiar one-dimensional Bloch band diagram (see figure 1). 
\begin{figure}[tbp]
\centerline{
\includegraphics[width=2.25in]{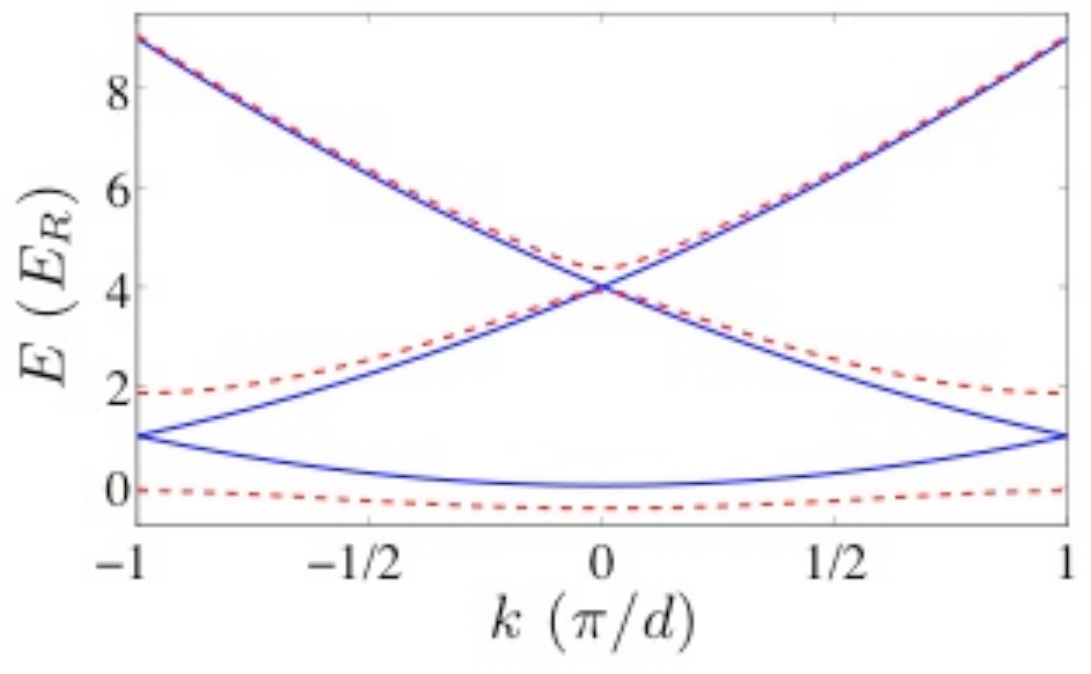}
}
\caption{Dispersion relation for free particles (solid) and particles moving in 
a one-dimensional sinusoidal lattice (dashed) plotted in a reduced zone scheme. 
Free space dispersion relations are plotted versus momentum, and lattice 
dispersion relations are plotted versus quasi-momentum. Energy is in units of the 
recoil energy, $E_R=\hbar^2\pi^2/2md^2$, where $d$ is the lattice spacing. The 
lattice depth is $V_0=5E_R$.}
\end{figure}

The introduction of a periodic potential breaks the continuous translational 
symmetry of the system, and therefore the wavefunctions are no longer momentum 
eigenstates. Instead, the eigenfunctions of
the Hamiltonian are eigenfunctions of a discrete translation
operator, $T_{d}$, that translates the entire system by one lattice spacing,
\begin{equation}
\psi \left( x-d\right) =T_{d}\psi \left( x\right) =e^{-ikd}\psi \left(
x\right),
\end{equation}%
and can be labeled by $k$, the \textit{quasi-momentum}. A quasi-momentum 
eigenstate of a periodic Hamiltonian can be written as the product of a 
plane wave and a function periodic in the lattice:
\begin{equation}
\psi^{(l)} _{k}\left( x\right) =e^{ikx}u^{(l)}_{k}\left( x\right),~~~~u^{(l)}_{k}\left(
x-d\right) =u^{(l)}_{k}\left( x\right),
\end{equation}
where $l$ is a band index.

When these Bloch functions are expanded in the
momentum basis, the only momenta contributing to the sum are those that
differ from $k$ by a reciprocal lattice vector, $G_j=\frac{2\pi j}{d}$. The expansion is 
then given by
\begin{equation}
\psi _{k}^{(l)}\left( x\right) =\int_{-\infty }^{\infty }\frac{dq}{\sqrt{2\pi }}%
e^{iqx}\psi _{k}^{(l)}\left( q\right) = \sum_{j=-\infty }^{\infty }c_j^{(l)} \frac{%
e^{i\left( k+2\pi j/d\right) x}}{\sqrt{2\pi }},
\end{equation}%
where the expansion coefficients $c_j^{(l)}$ are given by%
\begin{equation}
c_j^{(l)}=\int_{-\infty }^{\infty }dx\frac{e^{-i\left( k+2\pi j/d\right) x}}{%
\sqrt{2\pi }}\psi _{k}^{(l)}\left( x\right).
\end{equation}
Figure 2(a) is an example of a Bloch function in a sinusoidal lattice, and figure 2(b) is 
the Fourier transform of this Bloch function -- that is, the Bloch function in momentum space.
\begin{figure}
\centerline{
\includegraphics[width=4.5in]{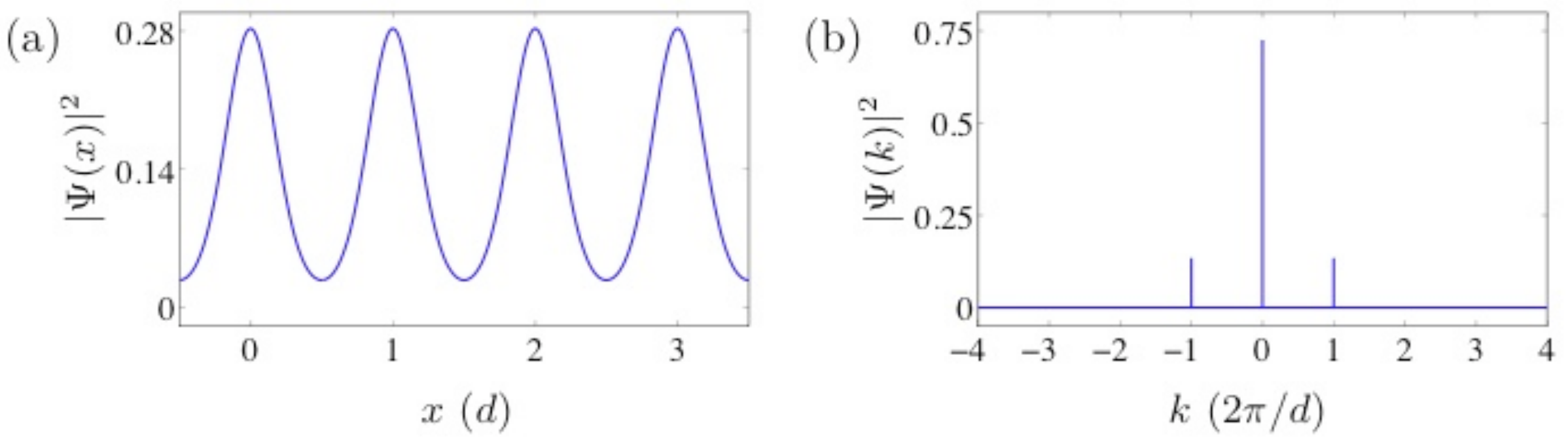}
}
\caption{(Color online.) The zero quasi-momentum Bloch function in the first band for an
infinite sinusoidal lattice if lattice depth $V_0=5E_R$. (a) The real space wave function is
concentrated in the lattice sites. (b) The wave function in momentum space
exhibits peaks at reciprocal lattice vectors. The envelope of these peaks is
given by a Gaussian for large lattice depths, consistent with a lattice of
simple harmonic oscillator wells.}
\end{figure}

In the rest of the paper, we will consider only lattices that are deep enough that 
a single-band approximation will be adequate to describe the physics. In the proceeding 
sections, then, we drop the band index, $l$.

\subsection{Lowering the symmetry}

If an extra potential, $V\left(x\right)$, satisfying
\begin{equation}
V\left( x-pd\right) =V\left( x\right),
\end{equation}
with $p\geq2$ an integer and $d$ the lattice spacing, is introduced to a lattice, then 
Bloch states $\left\langle x|k\right\rangle=\psi _{k}\left(x\right)$ and 
$\left\langle x|k^{\prime}\right\rangle=\psi _{k^{\prime}}\left(x\right)$ will mix with 
each other if $k$ and $k^{\prime }$ differ by $2\pi m/pd$, as shown by the following 
calculation:
\begin{equation}
\fl \left\langle k^{\prime }\left\vert V\right\vert k\right\rangle
=\int_{-\infty }^{\infty }dx\psi _{k^{\prime }}^{\ast}\left( x\right)
V\left( x\right) \psi _{k}\left( x\right) =e^{i\left(k^{\prime }-k\right)
pd}\left\langle k^{\prime }\left\vert V\right\vert k\right\rangle.
\label{mixeqn}
\end{equation}
Eigenstates of the new Hamiltonian will therefore be linear combinations of these Bloch 
states. The potential increases the lattice periodicity and, accordingly, 
decreases the size of the first Brillouin zone to $2\pi /pd$.

In figure 3(a), the lowest band of a one-dimensional lattice with an arbitrary 
periodic potential is plotted in a reduced zone scheme. Introducing an infinitesimal
potential whose periodicity is three times the lattice spacing $d$, the
first Brillouin zone is reduced to the region $-\pi /3d\leq k\leq \pi /3d$. 
Equation (\ref{mixeqn}) shows that the states lying on a vertical
line in figure 3(b) mix with each other under the addition of the new
potential -- i.e., $\left\langle k^{\prime }\left\vert V\right\vert k\right\rangle$ 
is non-zero for these states. In particular, the degenerate
states at $k=0$ mix, breaking the degeneracy and opening up a gap at $k=0$, 
as shown in figure 3(b). The single band splits into three bands due to
the reduction of the symmetry by $V\left(x\right) $.

When the addition of a potential reduces the symmetry of a system, as 
$V\left(x\right) $ did to the generic system described in figure 3, the
original eigenstates of the system mix according to their symmetry.
Accordingly, gaps open up where there are degeneracies between states 
of the same symmetry, creating a band structure.
\begin{figure}
\centerline{
\includegraphics[width=4.5in]{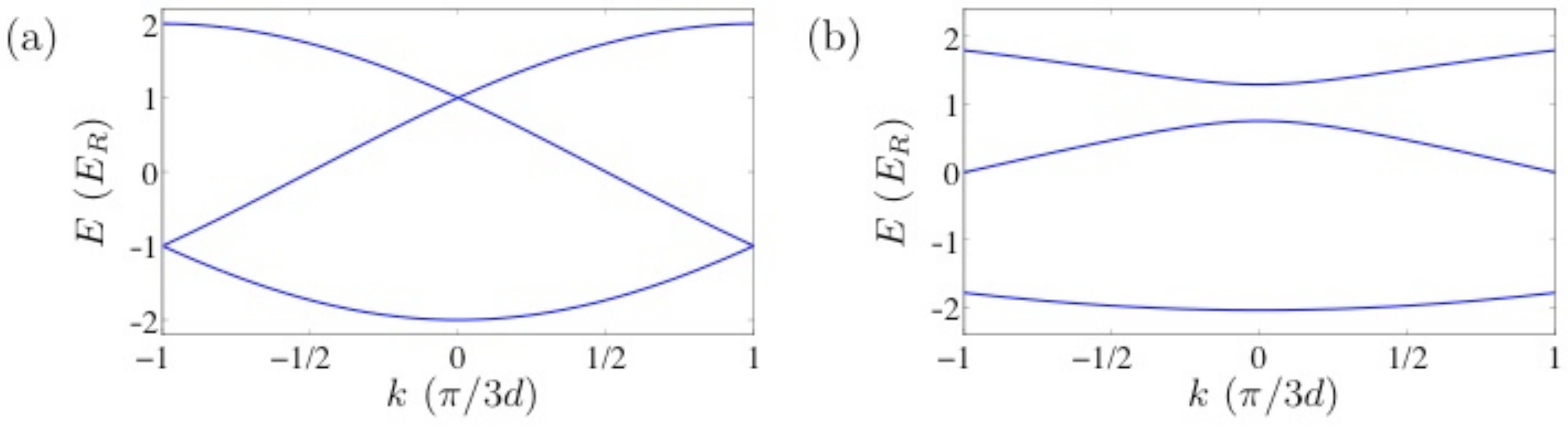}
}
\caption{(Color online.) (a) Lowest band dispersion relation for a one-dimensional 
sinusoidal lattice plotted in a reduced zone scheme that anticipates 
the lowering of the symmetry via an additional potential with periodicity $3d$. (b) After the
introduction of this potential, states with the same quasi-momentum, $k$ in the
reduced zone mix with each other, opening up gaps anywhere that there is a
degeneracy. The matrix elements of the added potential between states of the same 
quasi-momentum are of magnitude $5E_R$.}
\end{figure}

\subsection{The ground state of a moving lattice}

The utility of these symmetry concepts and quasi-momentum can be demonstrated 
by considering the effect of moving the lattice. In a stationary lattice, a zero 
quasi-momentum Bloch function is always the ground state. When
the lattice is moving, the energetically favored state is one that moves
along with the lattice. In order to investigate this phenomenon, we move to
a frame co-moving with the lattice. The Hamiltonian in this frame is given by 
\cite{landau1976m}
\begin{equation}
H=H_0-\vec{p}\cdot \vec{v},
\end{equation}%
where $H_0$ is the Hamiltonian in the non-moving frame, and $\vec{v}$ is
the velocity of the moving frame with respect to the lab frame. The ground state 
of $H_0-\vec{p}\cdot \vec{v}$ will be the ground state of the lab frame but 
written in rotating frame coordinates.

Consider an eigenstate of $H_0$, $\psi _{k}\left(
x\right) $. In
first-order perturbation theory, the energy of this state is%
\begin{equation}
E_{k}=\left\langle k\left\vert H_{0}-\vec{p}\cdot \vec{v}\right\vert
k\right\rangle =E^0_{k}-v\left\langle k\left\vert p\right\vert
k\right\rangle.
\end{equation}%
At zero velocity, the ground state is a zero quasi-momentum state, 
$\psi_0(x)$. However, the energy of a state that has a positive
average momentum, $\left\langle k\left\vert p\right\vert k\right\rangle
> 0$, falls below that of $\psi _{0}\left( x\right)$ for large enough $v$. 
This change is signalled by an exact energy level crossing 
in the energy spectrum between two quasi-momentum states (see figure 4). As 
the velocity is increased, the effect of the velocity term is to reorder the
energies of the quasi-momentum states, thereby leading to exact energy level
crossings.
\begin{figure}
\centerline{
\includegraphics[width=2.5in]{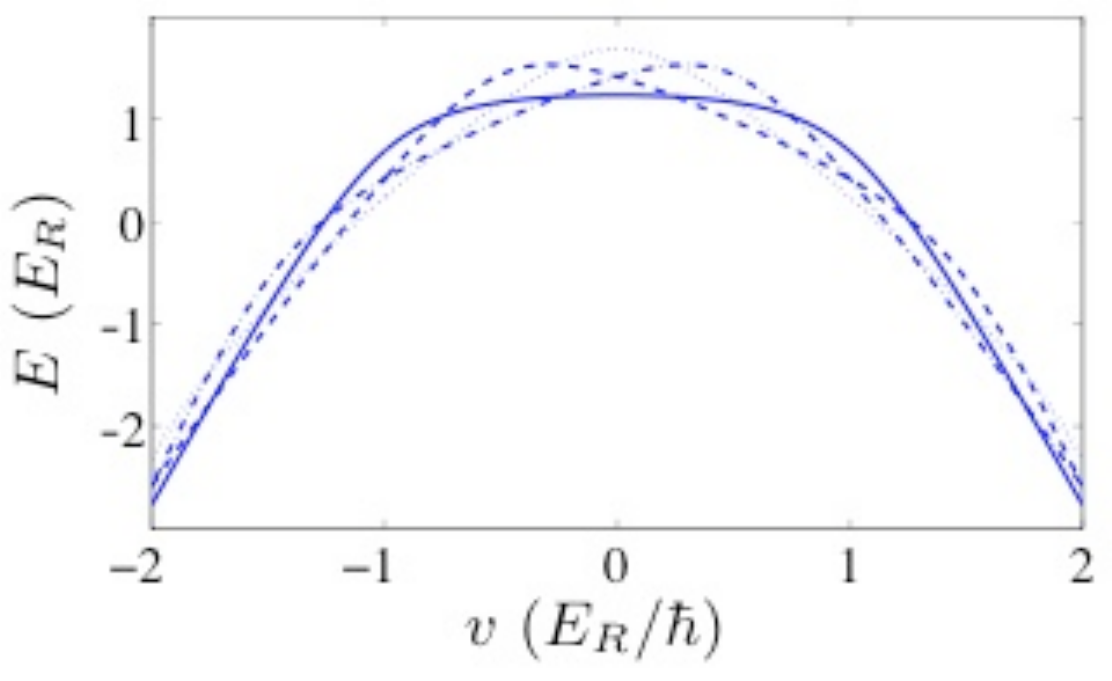}
}
\caption{(Color online.) The four lowest energies in the single-particle energy spectrum as
a function of lattice velocity for a four-site sinusoidal lattice with
periodic boundary conditions. At zero velocity, the zero quasi-momentum
state (solid) is the ground state. At some finite positive velocity, the 
$k=\pi /2d$ state (dashed) is energetically favorable and becomes the ground state. 
The ground state eventually cycles through all quasi-momenta.}
\end{figure}

\subsection{Quasi-angular momentum}

A moving $N$-site linear lattice with periodic boundary
conditions is analogous to a rotating $N$-site ring lattice. We can make this analogy 
explicit by considering the Hamiltonian of a moving, one-dimensional, 
sinusoidal $N$-site lattice in the co-moving frame, given by
\begin{equation}\label{H1D}
H=-\frac{\hbar ^{2}}{2M}\frac{\partial ^{2}}{\partial x^{2}}+V_{0}\cos
^{2}\left( qx\right)-v\frac{\hbar }{i}\frac{\partial }{\partial x},
\end{equation}
where $v$ is the velocity of the lattice. 
The wave-number $q$ can be rewritten as $q=\pi/d$, where $d$ is the lattice
spacing. If periodic boundary conditions,
\begin{equation}
\psi \left( x+Nd\right) =\psi \left( x\right),
\end{equation}%
are included, then the Hamiltonian describes the system illustrated in figure 5(a) 
for the explicit case of an 8-site lattice. On the other hand, the Hamiltonian of a 
rotating, sinusoidal $N$-site ring lattice (figure 5(b)) in the rotating frame is given by 
\begin{equation}\label{HRing}
H=-\frac{\hbar ^{2}}{2M}\frac{1}{R^{2}}\frac{\partial ^{2}}{\partial \phi
^{2}}+V_{0}\cos ^{2}\left( N\frac{\phi }{2}\right)-\Omega\frac{\hbar }{%
i}\frac{\partial }{\partial \phi },
\end{equation}
where $\Omega$ is the rotation frequency, $R$ is the radius of the ring, and 
$\frac{\hbar}{i}\frac{\partial}{\partial\phi}$ is the angular mometum operator, $L_z$. 
The inclusion of the term $-\Omega L_{z}$ has the effect of moving to a frame 
co-rotating with the lattice \cite{landau1976m}. The two Hamiltonians, equations 
\ref{H1D} and \ref{HRing}, are identical if we perform a transformation 
$x=\phi Nd/2\pi$ and identify $Nd/2\pi$ with $R$ and $v/R$ with $\Omega$.

Since the Hamiltonians are exactly identical, all of the properties of one-dimensional 
systems with a discrete translational invariance carry over for ring systems with a discrete 
rotational invariance. The analogy can be carried further, starting with the two-dimensional 
free-space solution in polar coordinates,
\begin{equation}
\psi _{j}\left( \phi ,\rho \right)=\e^{ij\phi }R_{j}\left( \rho \right),
\end{equation}%
where $R_{j}\left(\rho\right)$ is a radial function, irrelevant for our discussion, and 
$j$ is an integer. In the presence of a potential
that breaks the rotational symmetry, the eigenstates are linear combinations
of these free-space solutions. If this potential has a discrete $N$-fold
rotational symmetry, its eigenstates can be expanded in the free space
solutions as
\begin{equation}
\psi _{m}\left( \phi ,\rho \right) =\sum_{j=-\infty }^{\infty }
a_je^{i\left( Nj+m\right) \phi }R_{j}\left( \rho \right).
\end{equation}

It is evident from this expansion that $\psi _{m}\left( \phi
,\rho \right) $ is an eigenvector with eigenvalue $e^{-i2\pi m/N}$ of the discrete rotation operator 
$R_{2\pi /N}$ that rotates the system by the angle $2\pi/N$; 
$R_{2\pi /N}$ takes the place of the discrete translation 
operator, $T_{d}$. The analogy is complete when we note that
the eigenstates are linear combinations of angular momentum eigenstates, in
which case we call the number $\hbar m$ the quasi-angular momentum
of the state, $\psi _{m}\left( \phi ,\rho \right)$. For the rest of the
paper, we will drop the factor of $\hbar$ from the quasi-angular momentum.
\begin{figure}[tbp]
\centerline{
\includegraphics[width=3in]{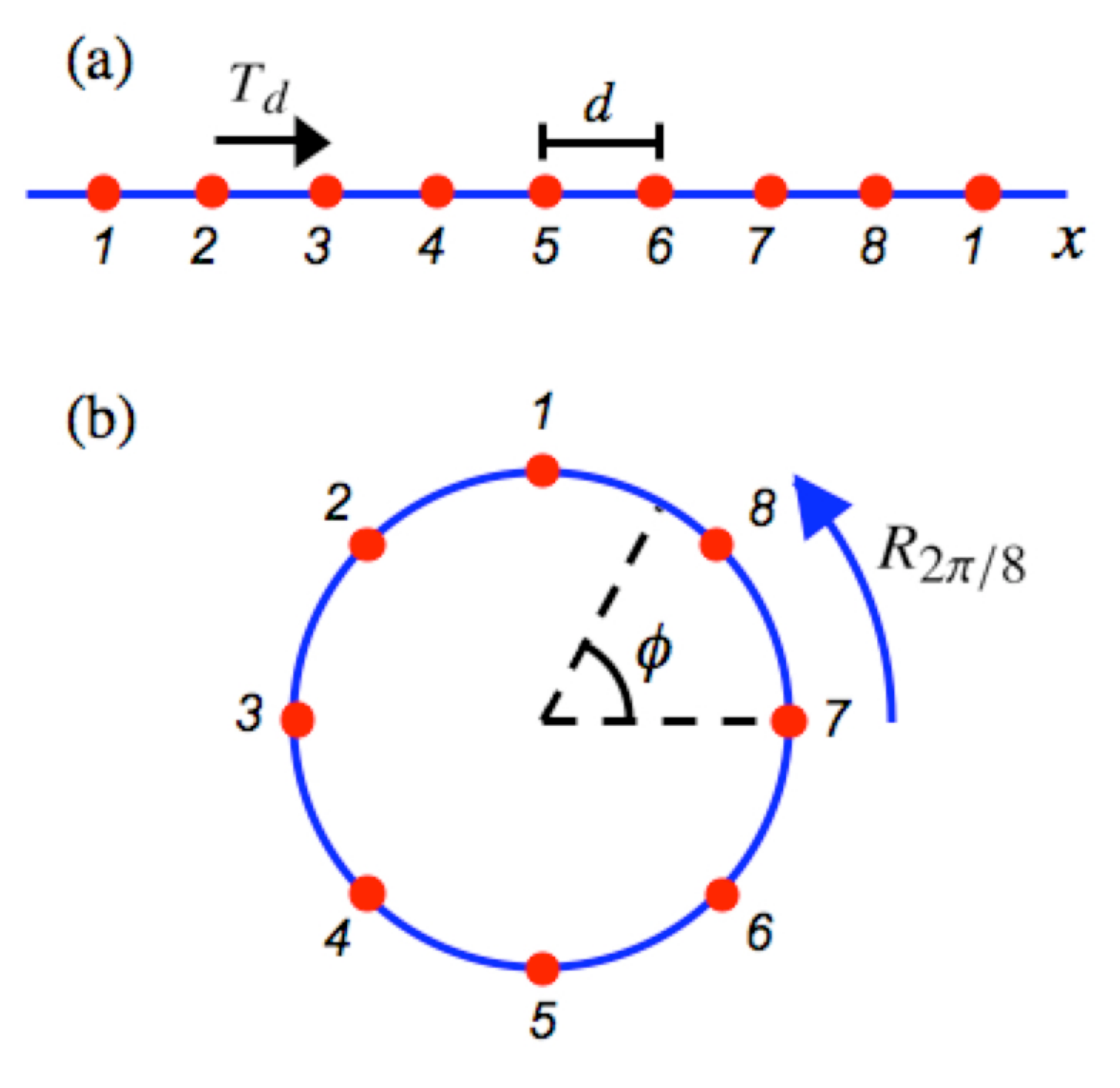}
}
\caption{(Color online.) This diagram shows the equivalence between a one-dimensional
lattice with periodic boundary conditions and a ring lattice with the same
number of sites. The translation operator $T_{d}$ that translates through
one lattice site maps onto the rotation operator $R_{2\pi/8}$ that
rotates the lattice through one lattice site.}
\end{figure}

\section{Bosons and fermions in rotating optical lattices}

Interacting bosons in one dimension enter the Tonks-Girardeau regime when 
interactions get very strong \cite{girardeau2004rbs,lieb1963eai}. 
The gas takes on some of the characteristics 
of a gas of non-interacting fermions, such as an identical number density. It differs 
in certain aspects, however, such as the momentum distribution. It then becomes 
particularly interesting to compare the cases of strongly-interacting bosons and 
non-interacting fermions in rotating lattices. In this section, gases of non-interacting 
fermions and hard-core bosons in rotating optical lattices are investigated. 
A many-particle formalism is generated for quasi-angular momentum states and 
used to monitor the symmetry of the ground state as a function of rotation.

The Hamiltonian for interacting bosons or fermions in a rotating optical
lattice takes the second quantized form, 
\begin{eqnarray}\fl
H =\sum_{\sigma }\int d^{D}x\hat{\Psi}_{\sigma }^{\dag }\left( \vec{x}%
\right) \left( -\frac{\hbar ^{2}}{2M}\nabla ^{2}+V_{lat}\left(\vec{x}\right)
-\Omega L_{z}\right) \hat{\Psi}_{\sigma }\left( \vec{x}%
\right)   \nonumber \\
\fl+\sum_{\lambda \lambda ^{\prime },\mu \mu ^{\prime }}\int d^{D}x\int
d^{D}x\hat{\Psi}_{\lambda }^{\prime \dag }\left( \vec{x}\right)\hat{\Psi}_{\lambda
^{\prime }}^{\dag }\left( \vec{x}^{\prime }\right) V\left( \vec{x},\vec{x}%
^{\prime }\right) _{\lambda \lambda ^{\prime },\mu \mu ^{\prime }}
\hat{\Psi}_{\mu^{\prime }}\left( \vec{x}\right) \hat{\Psi}_{\mu }\left( \vec{x}^{\prime }\right),
\end{eqnarray}%
where $D$ is the dimension of the system, $V_{lat}\left(\vec{x}\right)$ is taken to be sinusoidal, 
$\lambda$, $\mu$, and $\sigma$ 
are spin indices, and $\Psi _{\sigma }\left( \vec{x}\right) $, $\Psi _{\sigma }^{\dag
}\left( \vec{x}\right) $ are field operators satisfying the
(anti-)commutation relations,
\begin{eqnarray}
\left[ \hat{\Psi} _{\sigma }\left( \vec{x}\right) ,\hat{\Psi} _{\sigma ^{\prime }}^{\dag
}\left( \vec{x}^{\prime }\right) \right] _{\pm } &=&\delta _{\sigma \sigma
^{\prime }}\delta \left( \vec{x}-\vec{x}^{\prime }\right),   \nonumber \\
\left[ \hat{\Psi} _{\sigma }\left( \vec{x}\right) ,\hat{\Psi}_{\sigma ^{\prime }}\left( 
\vec{x}^{\prime }\right) \right] _{\pm } &=&0.
\end{eqnarray}
Assuming spin-independent contact interactions, a Hubbard
Hamiltonian can be derived in a lowest-band, tight-binding approximation 
by expanding the field operators in the lowest-band 
Wannier functions, $w(\vec{x}-\vec{x}_j)$ \cite{jaksch1998cba}. A basis better 
suited in the presence of rotation is given by the modified Wannier
functions,
\begin{equation}\label{modwan}
W_{j}\left( \vec{x}\right)
=\exp \left(-i\frac{M}{\hbar} \int^{\mathbf{x}}_{\mathbf{x}_j} \vec{A} \left( \vec{x}^{\prime} \right) \cdot d\vec{x}^{\prime} \right) w\left( \vec{x}-\vec{x}_{j}\right),
\end{equation}%
where $\vec{A}\left( \vec{x}^{\prime }\right) =\vec{\Omega}\times 
\vec{x}^{\prime }$ \cite{bhat2006qvs}. This modification captures the effect of 
rotation at low rotation speeds \cite{bhat2006qvs}; however, this approximation 
needs to be further modified at higher rotation speeds where the density of the 
Wannier functions is modified along with the phase.

The result for spinless bosons is \cite{bhat2007he}
\begin{eqnarray}
\fl H_{B} =-\sum_{\left\langle i,j\right\rangle }\left(  t+\frac{1}{2}M
\Omega^{2}  A_{1}\right)  \left(e^{-i\phi_{ij}}a_{i}^{\dag}%
a_{j}+e^{i\phi_{ij}}a_{j}^{\dag}a_{i}\right)\nonumber\\
+\sum_{j}\left(
\epsilon-\frac{1}{2}M\Omega^{2}  \left(
r_{j}^{2}+A_{2}\right)  \right)  a_{j}^{\dag}a_{j}+\frac{U}{2}\sum_{j}%
a_{j}^{\dag}a_{j}^{\dag}a_{j}a_{j}.
\end{eqnarray}
A similar derivation for spin-$1/2$ fermions yields%
\begin{eqnarray}
\fl H_{F} =-\sum_{\left\langle i,j\right\rangle ,\sigma}\left(  t+\frac{1}%
{2}M\Omega^{2}  A_{1}\right)  \left(e^{-i\phi_{ij}}
a_{i,\sigma}^{\dag}a_{j,\sigma}+e^{i\phi_{ij}}a_{j,\sigma}^{\dag}a_{i,\sigma
}\right)\nonumber\\
\fl +\sum_{j,\sigma}\left(  \epsilon-\frac{1}{2}M
\Omega^{2} \left(  r_{j}^{2}+A_{2}\right)  \right)
a_{j,\sigma}^{\dag}a_{j,\sigma}+\frac{U}{2}\sum_{j}a_{j,\uparrow}^{\dag
}a_{j,\downarrow}^{\dag}a_{j,\downarrow}a_{j,\uparrow}.
\end{eqnarray}%
For both cases, $i,j$ are site indices with $\left\langle i,j\right\rangle $ indicating a
sum over only nearest neighbours, and $r_i$ is 
the radial position of the $i$'th site. The phase $\phi_{ij}$ is given by the expression,
\begin{equation}
\phi_{ij}=\frac{M}{\hbar} \int^{\mathbf{x}_i}_{\mathbf{x}_j} \vec{A} \left( \vec{x}^{\prime}\right) \cdot d\vec{x}^{\prime}
=\frac{M\Omega }{\hbar}(x_i y_j-x_j y_i).
\end{equation}
The parameters $\epsilon$, $t$, and $U$ are the on-site zero-point energy, the hopping 
parameter, and the on-site interaction energy, respectively, identical to those of the 
standard Bose-Hubbard Hamiltonian \cite{jaksch1998cba}. They are given by the expressions,
\begin{eqnarray}
t \! &=& \! \int d^Dx ~ w^* (\vec{x}-\vec{x}_i) \left(-\frac{\hbar^2}{2M} \nabla^2 + V_{lat} \left(\vec{x}\right) \right) w(\vec{x}-\vec{x}_j), \\
\epsilon \! &=& \! \int d^Dx ~ w^* (\vec{x}-\vec{x}_i) \left(-\frac{\hbar^2}{2M} \nabla^2 + V_{lat} \left(\vec{x}\right) \right) w(\vec{x}-\vec{x}_i), \\
U&=&g \int d^Dx \left|  w(\vec{x}-\vec{x}_i)\right|^4,
\end{eqnarray}
where $g$ is a two-particle interaction strength. The parameters $A_1$ and $A_2$ arise due to 
the phase factor in equation \ref{modwan} and are given by
\begin{eqnarray}
A_1&=& \int dx ~ w^*\left(x-x_i\right) \left(x-x_i\right)^2 w\left(x-x_j\right),\\
A_2 &=& 2\int dx ~ w^*(x-x_i) \left( x-x_i \right) ^2 w\left(x-x_i\right),
\end{eqnarray}
where $w\left(x-x_i\right)$ are one-dimensional Wannier functions. All of these parameters 
can be numerically evaluated for a lattice of specific shape, period, and depth 
\cite{bhat2007he}. If a harmonic trap is included, 
$\Omega ^{2}\rightarrow \Omega^{2}-\Omega _{T}^{2}$, where $\Omega _{T}$ 
is the trap frequency.

In the rest of the paper, a lattice depth of $V_0=10E_R$ is assumed, and all 
parameters in the Hubbard Hamiltonians are numerically computed for this depth.

\subsection{Strongly-interacting bosons}

In this section, we describe hard-core bosons in a rotating optical lattice. The
quasi-angular momentum of an eigenstate for a one-dimensional system is then
defined and monitored as a function of rotation speed. The analytic results herein
derived are consistent with a previous numerical treatment of a two-dimensional system 
\cite{bhat2006qvs}.

In the limit of very strong interactions, $U\to\infty$, a gas of bosons enters the 
hard-core boson regime. In an optical lattice, this regime 
can be characterized by using a number basis where the occupation number of each site 
is either $l$ or $l+1$, with $l$ an integer. We can encode this 
fact in the Hamiltonian by formally changing the properties of the creation and 
annihilation operators. Here we consider the case of filling factors less than one, 
as the results for higher filling factors are qualitatively identical 
(see the appendix for the general case). This can be effected by
stipulating on-site anti-commutation relations for the operators -- i.e.,
\begin{equation}
\left[ a_{i},a_{i}^{\dag }\right]_+ =1,~~~~\left[ a_{i},a_{j\neq i}\right]_-=%
\left[ a_{i},a_{j\neq i}^{\dag }\right]_- =0.
\end{equation}
In this case, the interaction term vanishes, although strong-interactions still 
implicitly exist in the Hamiltonian. For practical purposes, $U=100t$, where
$t$ is the tunneling energy, is large enough to enter this regime \cite{bhat2006qvs}.

The Hamiltonian for a ring-lattice is given by
\begin{eqnarray}\label{1DBHH}\fl
H  =-\sum_{j=1}^{N}\left(  t+\frac{M\Omega^{2}}{2}
A_{1}\right)e^{i\Omega\Phi}a_{j+1}^{\dag}a_{j}+h.c.\nonumber\\
+\sum_{j=1}^{N}\left(  \epsilon-\frac{M\Omega^{2}}{2}M\Omega^{2}
\left(  R^{2}+A_{2}\right)  \right)  a_{j}^{\dag}a_{j},
\end{eqnarray}
where $R$ is the radius of the ring. The parameter $\Phi$ is given by
\begin{equation}
\Phi=\frac{M}{\hbar}(x_i y_j-x_j y_i),
\end{equation}
which is a constant on a ring. This Hamiltonian can be diagonalized by first performing the Jordan-Wigner transformation \cite{jordan1928upa,schultz1963ipb},
\begin{equation}
b_{j}=a_{j}e^{i\pi \sum_{i=1}^{j-1}a_i^{\dag }a_i},
\end{equation}%
which transforms the mixed $a$ operators into fully fermionic ones. The
transformed Hamiltonian takes different forms for even and odd numbers of
particles. Performing the canonical transformation,
\begin{equation}
d_{m}=\frac{1}{\sqrt{N}}\sum_{j=1}^{N}e^{-i\pi f_{m}j/N}b_{j},
\end{equation}%
where $f_{m}=2m$ for $n$ odd, $f_{m}=2m-1$ for $n$ even, and $n$ is the number 
of particles, results in 
\begin{eqnarray}
\label{egysp1} \fl H=\sum_{m=1}^{N}E_m d_{m}^{\dag}d_{m},\\
\label{egysp} \fl E_m=-2\left(  t+\frac{1}{2}M\Omega^{2}
A_{1}\right)  \cos\left(  \frac{\pi f_{m}}{N}-\Omega\Phi\right)
+\epsilon-\frac{1}{2}M\Omega^{2} \left(  R^{2}+A_{2}\right)
\end{eqnarray}%
The eigenstates are of the form $d_{m_{1}}^{\dag}\dots d_{m_{n}}^{\dag }
\left\vert 0\right\rangle$.

A complete picture of this diagonalization requires an interpretation of the 
eigenstates in terms of quasi-angular momentum. The discrete rotation operator 
is defined via
\begin{equation}
Rb_{j\neq N}^{\dag }R^{-1}=b_{j+1}^{\dag },~~~~Rb_{N}^{\dag }R^{-1}=\left(
-1\right) ^{n}b_{1}^{\dag },~~~~R\left\vert 0\right\rangle =\left\vert
0\right\rangle .
\end{equation}%
This operator commutes with the Hamiltonian, and its
action on an eigenstate, $\left\vert \psi \right\rangle =d_{m_{1}}^{\dag
}\dots d_{m_{n}}^{\dag }\left\vert 0\right\rangle $, is%
\begin{equation}
R\left\vert \psi \right\rangle =e^{-i\pi \left( f_{m_{1}}+\cdots
+f_{m_{n}}\right) /N}\left\vert \psi \right\rangle .
\end{equation}%
Therefore, the quasi-angular momentum of this state is
$\left(\frac{f_{m_{1}}+\cdots +f_{m_{n}}}{2}\right) \mod~N$.
A pictorial representation of this discussion is shown in figure 6 for a specific case 
with one particle.
\begin{figure}[tbp]
\centerline{
\includegraphics[width=4in]{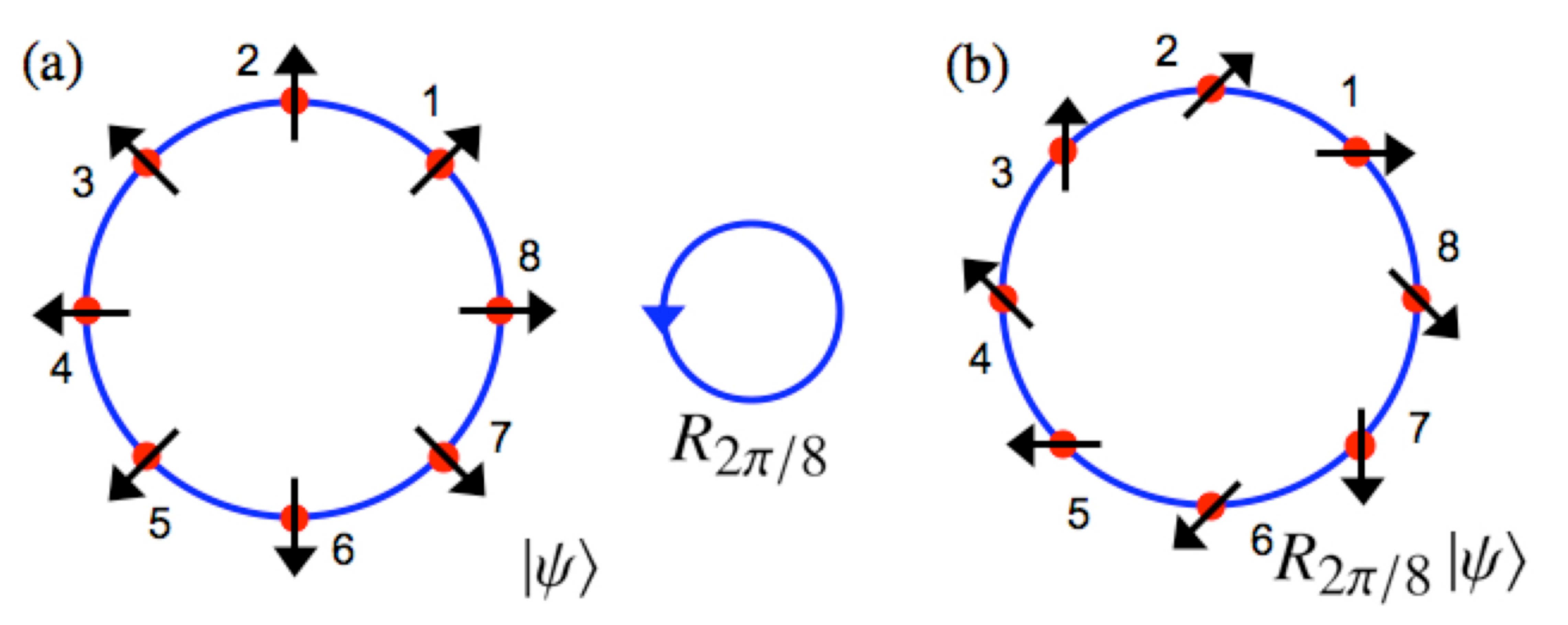}
}
\caption{(Color online.) (a) The operator $d_{1}^{\dag }$ adds one particle to the system,
populating every site with equal probability but a phase that differs by $2%
\pi /8$ between lattice sites. The state that is illustrated is $%
\left\vert\psi \right\rangle =d_{1}^{\dag }\left\vert
0\right\rangle $; arrows indicate phase. (b) When this state is rotated by
one lattice site, the resulting state is $R_{2\pi /8}\left\vert 
\psi \right\rangle $. Subtracting the phases at a site before and
after rotation, we get $\phi=-2\pi /8$. The picture then
illustrates the fact that $\left\vert\psi\right\rangle$ is an
eigenstate of $R_{2\pi /8}$ with eigenvalue $e^{-i2\protect\pi /8}$. }
\end{figure}

We are now in a position to use this formalism to investigate the ground state of the system. 
The ground state is determined by populating the lowest energy levels in the standard 
way for fermions. Equation \ref{egysp} is the single-particle energy spectrum $E_m$. 
As $\Omega$ is increased, the minimum value of the cosine moves to the 
right, as illustrated in figure 7(a). The ground state is then determined by populating 
the lowest energy $d_{m}^{\dag }d_{m}$ eigenstates. 
In figures 7(b) and 7(c), this process is illustrated for 
the case of three particles in an eight-site ring.
\begin{figure}
\centerline{
\includegraphics[width=4.5in]{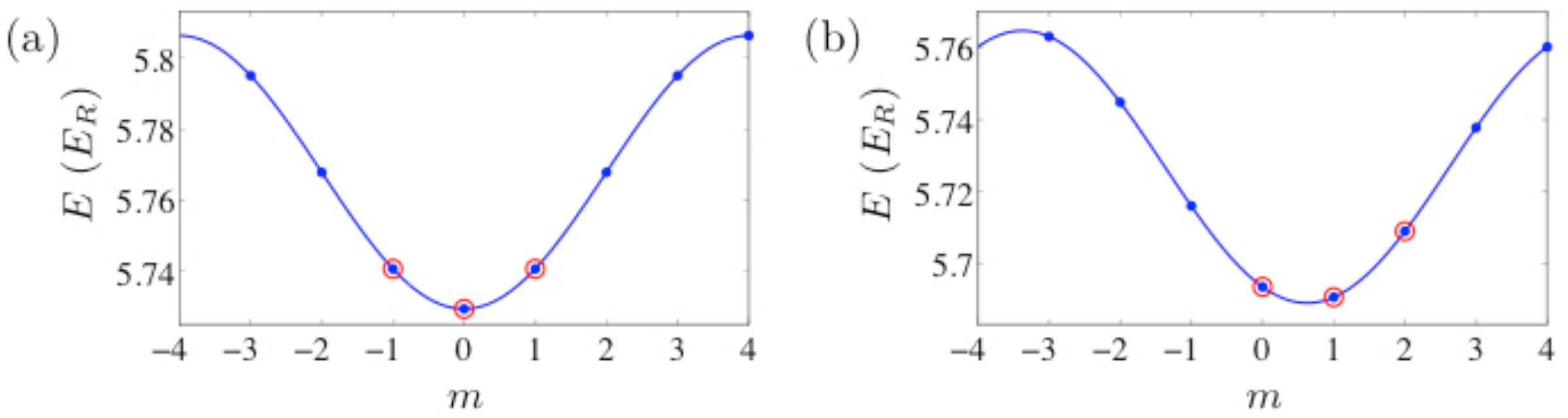}
}
\caption{(Color online.) Ground state of three strongly-interacting bosons in a rotating 
8-site ring lattice. Solid curves are guides for the eye. (a) At $\Omega=0$, the three 
lowest single-particle energies carry 
quasi-angular momentum $m=-1,0,1$. The three-particle ground state therefore
has quasi-angular momentum $m=\left( -1+0+1\right)\mod~8=0$. (c) At 
$\Omega=0.14E_R/\hbar$, the three lowest single-particle energies carry quasi-angular
momentum $m=0,1,2$. The three-particle ground state therefore has
quasi-angular momentum $m=\left( 0+1+2\right) \mod~8=3$.}
\end{figure}

In general, if the minimum is near $m=m_0$, we populate states
$m_0,~m_0\pm1,\dots,~m_0\pm p$ for $n=2p+1$ particles in the system. The 
quasi-angular momentum of the state is then
\begin{equation}
\fl m=\sum_{l=-p}^p \off{m_0+l}~\mod~N=m_0\off{2p+1}~\mod~N=m_0n~\mod~N.
\end{equation}
A similar argument holds for even numbers of particles. Thus, as rotation is 
increased from $\Omega=0$, the quasi-angular momentum cycles through the values,
\begin{equation}
\label{mb} m=jn~\mod~N:j=0,1,2,\dots.
\end{equation}

One final note for the case of bosons. If there is an infinitesimal
potential, $V$, that breaks the symmetry of the original Hamiltonian, $H_0$, 
down to a four-fold rotational symmetry, the behavior described in section 2.2 obtains. 
We treat the problem perturbatively. Tiny bandgaps in the single-particle energy 
spectrum, equation \ref{egysp}, open up at the new Brillouin zone boundaries. The
eigenstates of $H_0$ are independent of $\Omega$. Therefore, to first
order in $V$, the energy level shifts are independent of $\Omega$, and the
eigenstates remain unchanged. The effect of rotation is then exactly the
same as in the non-perturbed case, and the quasi-angular momenta have the
same behavior as a function of rotation. However, since the old
quasi-angular momentum is no longer a good quantum number, we have to take
\begin{equation}
m\to m~\mod~4.
\end{equation}
These results are consistent with the treatment in a previous paper \cite{bhat2006qvs}.

\subsection{Non-interacting fermions in a ring lattice}

Previous results focused on strongly-interacting bosons in rotating optical lattices 
\cite{bhat2006bec,bhat2006qvs}. Here, we describe non-interacting fermions 
in ring lattices and compare to the boson case. The quasi-angular 
momentum of the ground state changes as rotation is increased. The allowed 
values of this quantity for fermions differ from those of the boson case.

For non-interacting fermions, we drop the spin index, and the Hamiltonian for 
a ring lattice is identical to equation \ref{1DBHH} except that the operators 
are fermionic. This Hamiltonian can be analytically diagonalized via the 
canonical transformation 
\begin{equation}
d_{m}=\frac{1}{\sqrt{N}}\sum_{j=1}^{N}e^{-i2\pi mj/N}a_j.
\end{equation}
The Hamiltonian is identical to that of equations \ref{egysp1} and \ref{egysp} except that 
$f_m=2m$ for both even and odd numbers of particles.
The eigenstates, $d_{m_{1}}^{\dag }\cdots d_{m_{n}}^{\dag
}\left\vert 0\right\rangle $, carry quasi-angular momentum $m=\left(
m_{1}+\cdots +m_{n}\right) \mathrm{mod}N$, where $n$ is the number of particles. 
A discussion similar to the one given for bosons yields $m$ as a function of rotation:
\begin{eqnarray}
m=nj~\mod~N;~~~~j=0,1,2,\dots;~~~~n~odd,\\
\label{mfeven} m=n\left( j+\frac{1}{2}\right)~\mod~N;~~~~j=0,1,2,\dots;~~~~n~even.
\end{eqnarray}

As expected, the single-particle cases for both bosons and fermions are identical. 
It turns out that the cases of odd numbers of particles also coincide.
However, there is an interesting distinction between the two cases when the number 
of particles in the system is even (compare equations \ref{mb} and \ref{mfeven}). 
For instance, a system of two fermions in a four-site 
lattice cycles between quasi-angular momenta $1$ and $3$ whereas a system of two 
bosons cycles between values $2$ and $4$. This is one way in which non-interacting 
fermions and strongly-interacting bosons in one dimension differ despite the 
Jordan-Wigner transformation that maps between the two cases.

\subsection{Non-interacting fermions in a two-dimensional square lattice}

Differences between fermions and bosons remain when considering two-dimensional 
systems. Here we describe non-interacting fermions in a two-dimensional $4\times4$ 
square lattice.

The Hamiltonian can 
be written as the sum of three terms, $H=H_{12}+H_{4}+V$, where $H_{12}$ 
is the Hamiltonian of the outer twelve-site ring, $H_{4}$ is the Hamiltonian of 
the inner four-site ring, and $V$ is the hopping between the rings,
\begin{eqnarray}
V=-\sum_{\left\langle i,j\right\rangle}\left(t+\frac{M\Omega^{2}}{2}A_{1}\right)
a_{i}^{(12)\dag}a_{j}^{(4)}e^{-i\phi_{ij}}+h.c.,
\end{eqnarray}
where the superscripts $4$ and $12$ indicate whether site $j$ is on the outer 
$12$-site or inner $4$-site ring.
The Hamiltonians $H_{4}$ and $H_{12}$ can be separately diagonalized 
(apart from the term in $H_{12}$ involving $r_i^2$) via the transformations,
\begin{equation}
d_{m}^{\left( N\right) }=\frac{1}{\sqrt{N}}\sum_{j=1}^{N}e^{-i2\pi
mj/N}a_j^{\left( N\right) }.
\end{equation}

In figure 8(a) is plotted the single-particle energy spectrum for the
Hamiltonian in the absence of hopping between the rings ($V=0$). When the
hopping is ``turned on'' (figure 8(b)), level crossings become avoided
crossing due to the mixing of states with the same four-fold quasi-angular
momentum. In figure 8(c), the case for zero hopping between rings is plotted for 
$\Omega=0.05E_R/\hbar$, whereas in figure 8(d), the hopping has turned
on, and the avoided crossing behavior is observed.
\begin{figure}
\centerline{
\includegraphics[width=4.5in]{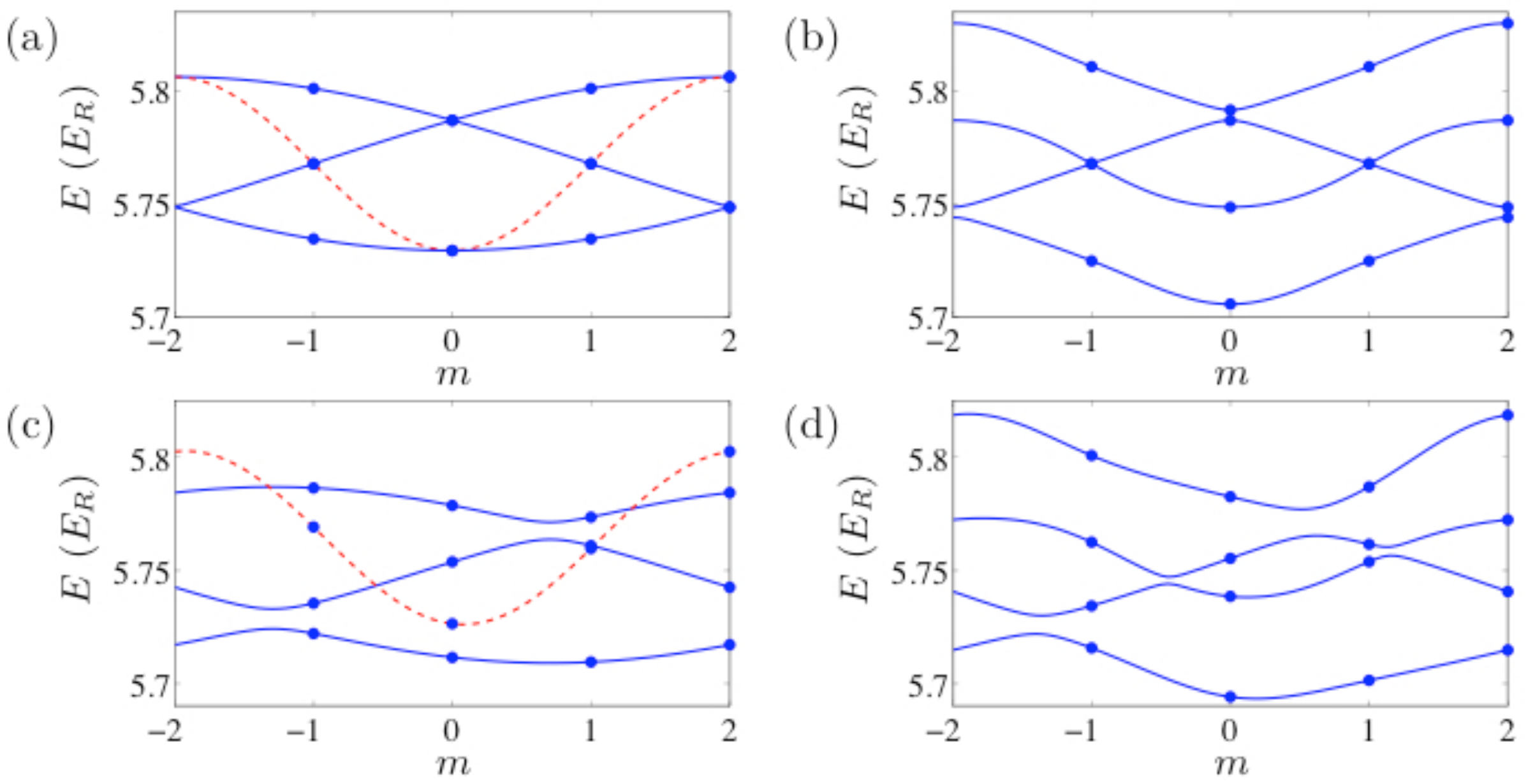}
}
\caption{(Color online.) Single-particle energy spectrum in a $4\times4$ square 
lattice for (a) zero rotation and zero hopping between inner and outer rings, 
(b) zero rotation and non-zero 
hopping, (c) $\Omega=0.05E_R/\hbar$ and zero hopping, and (d) 
$\Omega=0.05E_R/\hbar$ and non-zero hopping between inner and outer rings. 
The dashed and solid line are aids to the eye that indicate the band. In (a) and (c), 
the red dashed line is the spectrum for the inner 4-site lattice.}
\end{figure}

In figure 9 are plotted the quasi-angular momenta of the ground state as a 
function of rotation speed for $1$, $2$, $3$, $4$, and $5$ non-interacting 
fermions in a sixteen-site lattice. Similar to the ring lattice case, the results for 
odd numbers of particles are identical to those for hard-core bosons, but differ 
for even numbers of particles. However, there is an extra distinction that does not 
occur in the ring lattice. The pattern of $m$ values taken on for five particles 
(see figure 9(e)) differs markedly from that of one particle (see figure 9(a)), whereas 
for bosons these patterns are essentially identical \cite{bhat2006qvs}.
\begin{figure}
\centerline{
\includegraphics[width=4.2in]{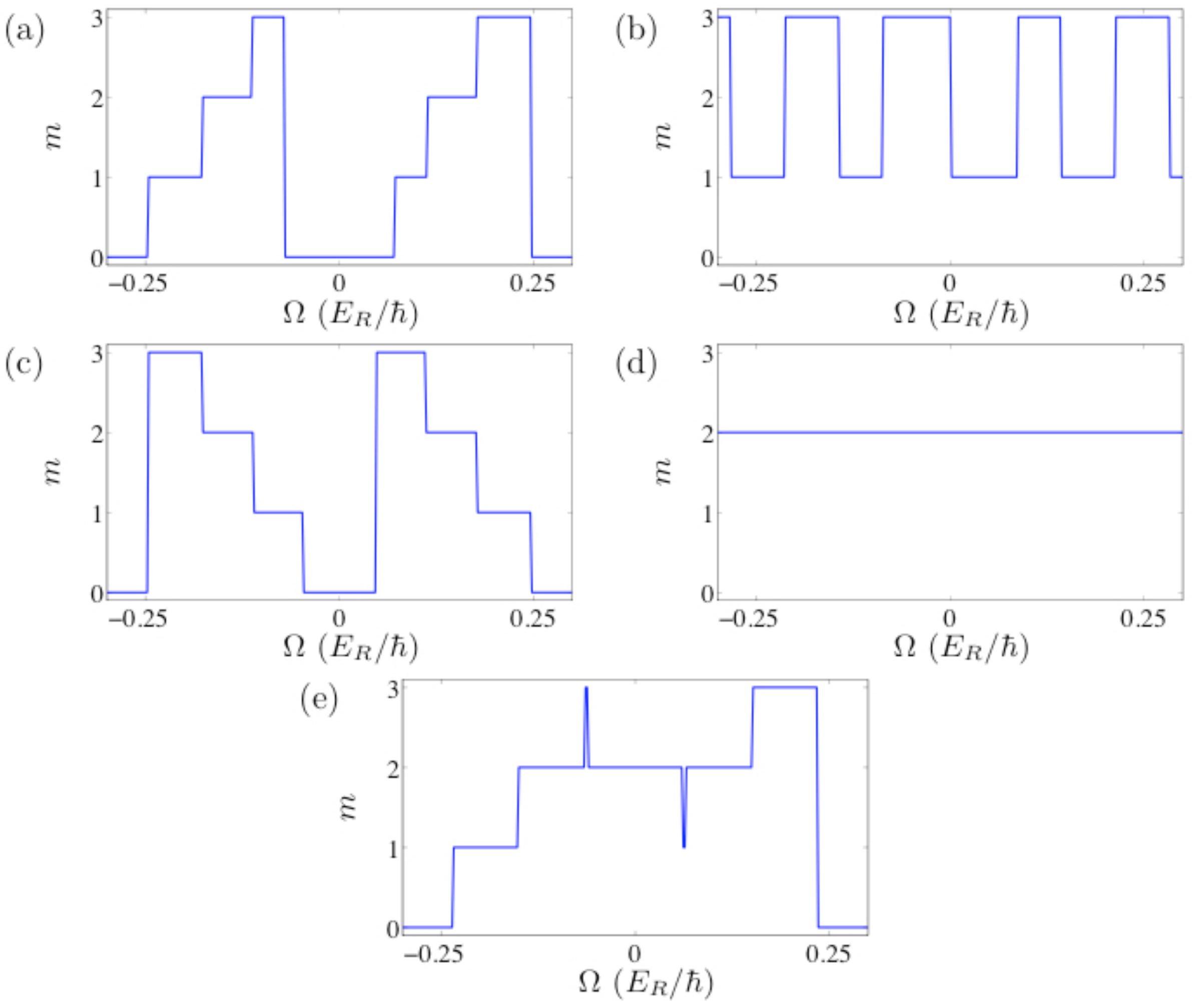}
}
\caption{(Color online.) Quasi-angular momentum as a function of rotation for (a) 1, (b) 2,
(c) 3, (d) 4, and (e) 5 non-interacting fermions in a sixteen site square
lattice. These patterns repeat for larger rotation rates.}
\end{figure}

\section{Signatures of quasi-angular momentum in the momentum distribution}

While the quasi-angular momentum $m$ is a good quantum number and a useful
tool in investigating the symmetry properties of the ground state, it is not
a quantity that can be directly measured in experiments. Instead, experiments are 
routinely performed in which time-of-flight, far-field, density images of the gas 
contain directly information about the momentum distribution before expansion. As a first 
step towards detecting quasi-angular momentum states, we look for a signature of 
quasi-angular momentum in the momentum distribution of a state.

The momentum distribution of the many-body state, $\ket{\psi}$, is given by
\begin{equation}
n\of{\vec{k}}=\bra{\psi} \hat{\Psi}^{\dag}(\vec{k}) \hat{\Psi}\of{\vec{k}} \ket{\psi},
\end{equation}
where $\hat{\Psi}(\vec k)$ is the Fourier transform of the field operator $\hat{\Psi}(\vec x)$. 
Recalling that we expand the field operators as
\begin{equation}
\hat{\Psi}(\vec{x})=\sum_j a_j W_j(\vec{x})=\sum_j a_j e^{-i\phi_{ij}}w\left( \vec{%
x}-\vec{x}_{j}\right) ,
\end{equation}
it can be shown that the momentum distribution takes the form
\begin{equation}\label{momdist}
n\of{\vec{k}}=\sum_{ll^{\prime }}w^{\ast }\of{\vec{k}_{l}}
w\of{\vec{k}_{l^{\prime }}} e^{i\vec{k}\cdot \left( \vec{x}_{l}-\vec{x}_{l^{\prime }}\right) }
\left\langle\psi \right\vert a_{l}^{\dag }a_{l^{\prime }}\left\vert \psi \right\rangle,
\end{equation}
where
\begin{equation}\label{kl}
\vec{k}_l=\vec{k}+\Omega\frac{M}{\hbar}\vec{x}_l\times\vec{\hat{z}}.
\end{equation}%
This distribution is written in the momentum coordinates, $\vec{k}$, corresponding 
to the rotating frame coordinates, $\vec{x}$. Since the number density rotates in the lab 
frame coordinates, $\vec{x}_L$, the momentum distribution also rotates in the lab frame 
momentum coordinates, $\vec{k}_L$. However, far-field pictures of the gas will be 
snapshots of the momentum distribution at the moment the trap and lattice are turned off, 
provided the switch-off time is fast enough. Thus, the momentum distributions presented 
here are accurate representations of what will be measured in time-of-flight measurements, 
although they may be rotated relative to each other.

Since quasi-angular momentum is a reflection of the symmetry of the system, considering 
single-particle states should be sufficient to capture the basics of how this quantum 
number affects the momentum distribution. We therefore concentrate on single-particle 
states only. In this case, $n\of{\vec{k}}$ is merely the square of the Fourier transform of 
the wavefunction.

In order to build up an understanding of the momentum distribution, first drop 
the Wannier functions. In this case, $n\of{\vec{k}}$ is the Fourier transform of a sum 
of weighted delta functions. On a four site-lattice, the wavefunction is of the form
\begin{equation}
\psi \left( \vec{x} \right)=\sum_{j=1}^4 e^{i2\pi mj/4} \delta \left( \vec{x}-\vec{x}_j \right),
\end{equation}
where $\vec{x}_j$ are the four corners of a square centered at $\vec{x}=0$. In figure 
10 are plotted the momentum distributions for $m=0$, $1$, and $2$. 
The $m=0$ is peaked at $\vec{k}=2\pi n/d$, with $n$ an integer and $d$ the lattice 
spacing, whereas the other 
states vanish at these points. The spacing between peaks for both $m=0$ and $m=2$
is $\Delta k=2\pi/d$, whereas for $m=1,3$ this spacing is $\Delta k=2\pi/%
\sqrt{2}d$; these can be calculated analytically. Thus, since the lattice spacing is known, 
there are clear measurable distinctions between different generic quasi-angular 
momentum states. Note that $m=1$ and $m=3$ are identical as they represent similar 
states with counter-propagating current patterns.
\begin{figure}[h]
\centerline{
\includegraphics[width=5.2in]{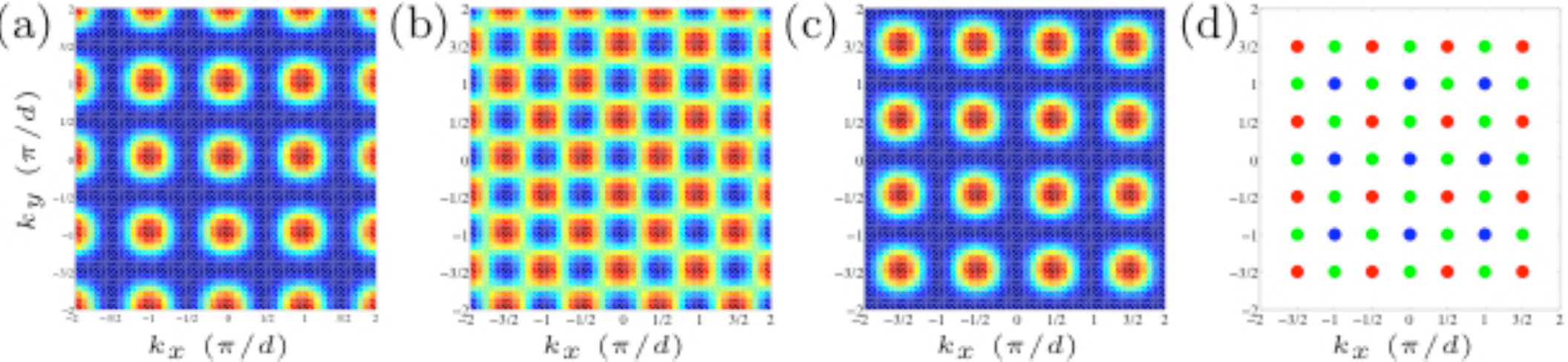}
} 
\caption{(Color online.) Fourier transform of generic (a) $m=0$, (b) $m=1,3$, and (c) $m=2$
quasi-angular momentum states on four sites. These momentum distributions
are plotted over a range $4\pi/d \leq k_x,k_y \leq 4\pi/d$;
the first Brillouin zone is given by $\pi/d \leq k_x,k_y \leq 
\pi/d$. (d) A schematic of the peak structure of the momentum
distribution of four delta functions of the $m=0$ (blue), $m=1,3$ (green),
and $m=2$ (red) quasi-angular momentum states. The peak spacings for the $m=0
$ and $m=2$ states are both $2\pi/d$, where $d$ is the spacing
between the delta functions, whereas the peak spacing for $m=1,3$ is $2
\pi/\sqrt{2}d$.}
\end{figure}

Re-including the Wannier functions contributes to the 
momentum distribution an overall envelope of approximate width $\sqrt{V/E_R}/4$, 
where $V$ is the depth of the lattice. The modification $\vec{k}\to\vec{k}_l$, equation 
\ref{kl}, moves the peak of this envelope away from $\vec{k}=0$ as rotation is increased, 
though the center of the envelope is always located at $\vec{k}=0$. This is consistent with 
higher momenta being accessed for higher rotation rates.

At this point, it is important to note that locating $\vec{k}=0$ in the distribution is necessary 
for distinguishing the $m=0$ and $m=2$ states. This can be done easily due to the overall
envelope in momentum space that is centered at $\vec{k}=0$.

There are two additional considerations that come into play for larger lattices, 
both system-size effects. An overall envelope in position space determines the width 
of the peaks in momentum space. This envelope determines the size of the system, 
and if the width is roughly $L$, then the width of the peaks in momentum space scales 
as $1/L$. If the lattice spacing is increased by a factor of $n$, then the peak-spacing in 
momentum space is decreased be a factor of $1/n$. The general peak structure is not 
modified by these considerations.

Since the peak structure is not modified by lattice-size effects, we can conclude that 
the distinctions between quasi-angular momentum states are sustained for more 
physical systems, which we turn to next. The ground state momentum distribution 
for a one-hundred-site lattice in the presence of a harmonic trap of frequency 
$\Omega_T = 0.15E_R/\hbar$ is computed via imaginary time propagation. The 
trap ensures that the majority of the number density resides on the inner 36 sites.
\begin{figure}[h]
\centerline{
\includegraphics[width=5in]{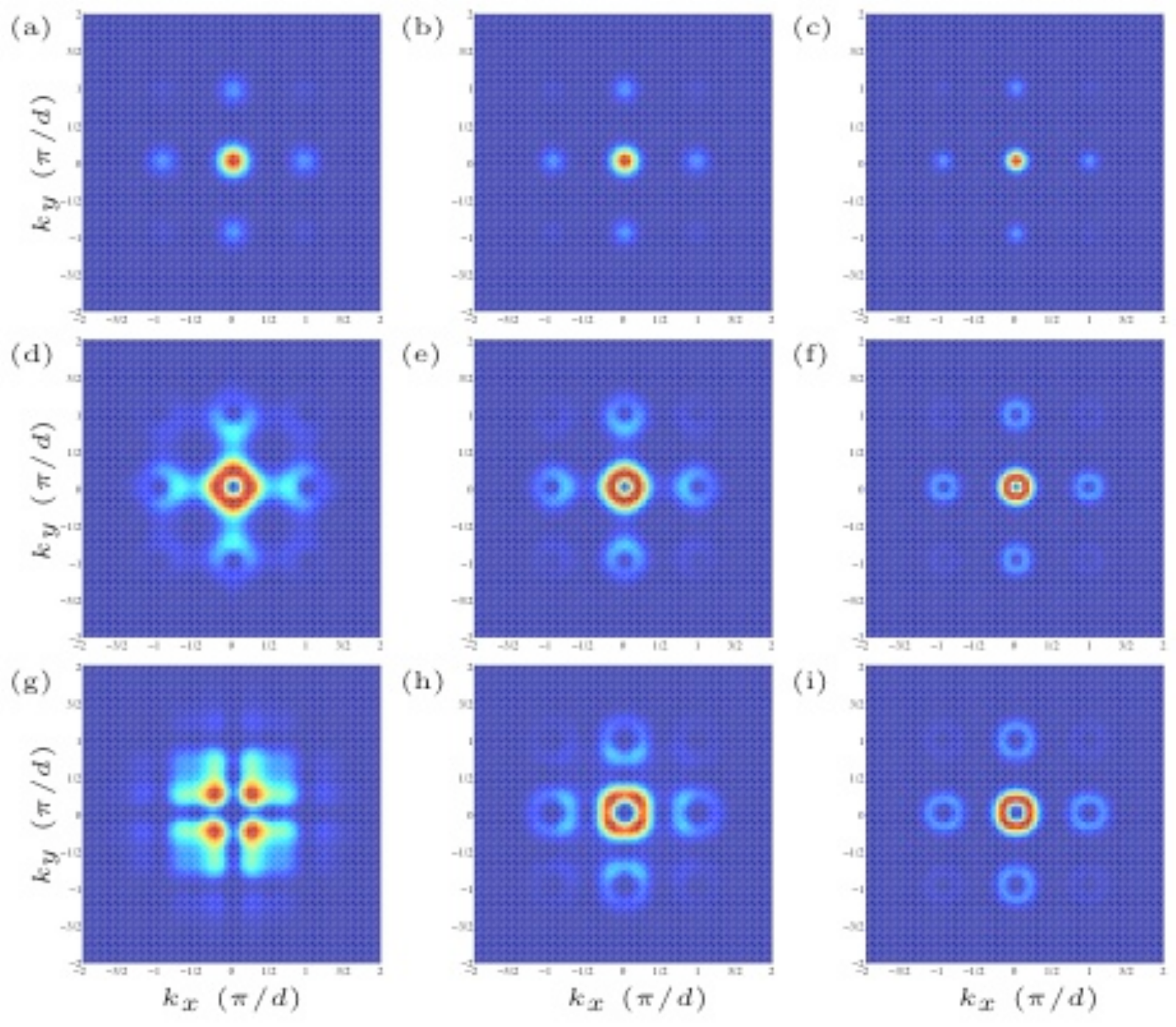}
}
\caption{(Color online.) Momentum distributions computed via imaginary time 
propagation for one particle in a 100-site lattice in the presence of a trap of 
frequency $\Omega_T=0.15E_R/\hbar$. Plotted 
are $m=0$ ((a), (b), and (c)), $m=1$ ((d), (e), and (f)), and $m=2$ ((g), (h), and (i)) 
states for rotation speeds of $\Omega=0$ ((a), (d), and (g)), $\Omega=0.1E_R/\hbar$ 
((b), (e), and (h)), and $\Omega=0.145E_R/\hbar$ ((c), (f), and (i)). At zero rotation, the 
distinctions between different quasi-angular momenta are sustained. As 
rotation is increased, it becomes harder to distinguish $m=1$ and $m=2$ 
states. Due to the centrifugal term, higher rotation increases the width of the 
envelope of the number density, decreasing the width of the peaks in 
momentum space. Not shown is $m=3$ which has a structure similar to both 
$m=1$ and $m=2$ for large rotation but has a ring-radius larger than both of them.}
\end{figure}

The results are summarized in figure 11, in which are plotted the momentum
distributions for $m=0$, $1$, and, $2$ states at rotation speeds of $0$, $0.1E_R/\hbar$, 
and $0.145E_R\hbar$, 
increasing to the right. We can again make the observation that the $m=0$ state 
is the only one that is non-zero at exactly the reciprocal lattice vectors, i.e. 
$\vec{k}=\frac{2\pi n}{d}\vec{\hat{k}}_x+\frac{2\pi m}{d}\vec{\hat{k}}_y$. The peak-spacing is 
smaller for the $m=1$ state than for the $m=2$ state, distinguishing them from each other. 
However, the exact peak-spacing is unclear due to complex interference effects between 
sites at different radii.

These differences are washed out at higher rotation speeds. The $m=1$, $m=2$, 
and $m=3$ (not pictured) states have very similar structures for large rotation. The peaks 
overlap so that it is very difficult to resolve them; the momentum distributions then all appear 
to be rings centered at reciprocal lattice vectors. They are distinguishable via the radius 
of the ring, as the radius is smallest for $m=1$ and largest for $m=3$, but as this is 
in order of increasing energy, the larger radius may just be an artifact of the larger 
width in the number density envelope at higher energies. Indeed, figures 11a, 11b, and 11c
are consistent with this conclusion, as the momentum peaks are narrowing due to 
the number density spreading out in real space. These results are consistent with 
those calculated via direct diagonalization of the Hubbard models previously discussed.

\section{Conclusion}

By analogy with quasi-momentum in translationally invariant periodic
systems, the notion of quasi-angular momentum was introduced to label the
eigenstates of a Hamiltonian that has a discrete rotational symmetry. It was
shown that quasi-angular momentum is useful in analyzing the
ground-state properties of quantum gases of bosons or fermions in rotating 
optical lattices. In particular, monitoring the quasi-angular momentum of the ground 
state as a function of rotation allowed us to identify transitions between different 
circulation values.

We also presented a possible avenue by which the quasi-angular
momentum of a state can be experimentally determined. We identified 
characteristics in the momentum distribution distinguishing between different
quasi-angular momentum states at low rotation speeds, such as the existence
of a peak at reciprocal lattice vectors for an $m=0$ state only, the peak-spacing, and the
overall structure of the momentum distribution.

There are still open questions as to how the momentum distributions will
change when the lattice size or number of particles is increased. The effects of 
statistics will be of fundamental importance for many-particle systems; there is a 
question of how well the peaks can be resolved for larger lattices, since peak-%
spacing decreases with increasing lattice size; and interference effects 
between adjacent sites and between rings is not rigorously treatable. However, 
there will always be signatures of the quasi-angular momentum in
the momentum distribution, as it is connected with the phase
information of the ground state, which in turn influences the momenta of
the system. In addition, a zero quasi-angular momentum 
state can be distinguished from a non-zero one, which allows the experimenter 
to verify that vorticity has entered the system.

\section*{Acknowledgements}

We acknowledge extremely useful discussions with both Brian Seaman and John
Cooper. The authors would also like to acknowledge funding support from the
Department of Energy, Office of Basic Energy Sciences via the Chemical
Sciences, Geosciences, and Biosciences Division, NASA, and Deutsche
Forschungsgemeinschaft (MK).

\appendix

\section*{Appendix}

\setcounter{section}{1}

The case of fillings between $l$ and $l+1$ can be encoded in the Hamiltonian
by formally changing the properties of the creation and annihilation
operators. As mentioned in the body of the paper, for fillings less than
one, the creation and annihilation operators satisfy the following
relations: 
\begin{equation}
\left\{ a_{i},a_{i}^{\dag }\right\} =1,~~~~\left[ a_{i},a_{j}\right] =\left[
a_{i},a_{j}^{\dag }\right] =0.
\end{equation}
If the filling fraction is between $l$ and $l+1$, the following changes are
made. First, the site number operator $a_{j}^{\dag }a_{j}$ is replaced by $%
a_{j}^{\dag }a_{j}+l$, reflecting the fact that each site is filled with
between $l$ and $l+1$ particles. Secondly, recalling that the operators were
originally bosonic, $a_{j}^{\dag }\left\vert l\right\rangle =\sqrt{l+1}%
\left\vert l+1\right\rangle $ and $a_{j}\left\vert l+1\right\rangle =\sqrt{%
l+1}\left\vert l\right\rangle $, one can see that the hopping parameter $t$
is scaled by a factor of $l+1$. Finally, the interaction term gets modified
according to the replacement $a_{j}^{\dag }a_{j}\rightarrow a_{j}^{\dag
}a_{j}+l$:
\begin{equation}
\sum_{j=1}^{N}a_{j}^{\dag }a_{j}\left( a_{j}^{\dag }a_{j}-1\right)
\rightarrow 2l\sum_{j=1}^{N}a_{j}^{\dag }a_{j}+Nl\left( l-1\right)
\end{equation}
The Hamiltonian in two-state approximation is then given by%
\begin{eqnarray}
\fl H =\left( Ul-\mu \right) \sum_{j=1}^{N}a_{j}^{\dag }a_{j}-\mu Nl+UN\frac{%
l\left( l-1\right) }{2}
-\left( l+1\right) \sum_{j=1}^{N}te^{i\phi}a_{j+1}^{\dag }a_{j}+h.c.
\end{eqnarray}
where the chemical potential $\mu$ has been introduced so that we can work
in the grand-canonical ensemble.

Upon applying the Jordan-Wigner transformation,
\begin{equation}
b_{j}=a_{j}e^{i\pi \sum_{k=1}^{j-1}a_{k}^{\dag }a_{k}},
\end{equation}
the Hamiltonian becomes
\begin{eqnarray}
\fl H=\left( Ul-\mu \right) \sum_{j=1}^{N}b_{j}^{\dag }b_{j}-\mu Nl+UN\frac{%
l\left( l-1\right) }{2}
-\left( l+1\right) \sum_{j=1}^{N-1}te^{i\phi}b_{j+1}^{\dag }b_{j}+h.c.  \nonumber \\
-\left( -1\right) ^{n+1}\left( l+1\right)te^{i\phi}b_{1}^{\dag }b_{N}+h.c.
\end{eqnarray}%
where $n$ is the number of particles. Now that the Hamiltonian is fully
fermionic, we can perform a canonical transformation to diagonalize it,%
\begin{equation}
d_{m}=\frac{1}{\sqrt{N}}\sum_{k=1}^{N}e^{i\pi f_{m}k/N}b_{k},
\end{equation}%
where $f_{m}=2m$ for $n$ odd and $f_{m}=2m-1$ for $n$ even. The resulting
Hamiltonian is 
\begin{eqnarray}
H =-\mu Nl+\frac{U}{2}Nl\left( l-1\right) +\sum_{m=1}^{N}E_{m}d_{m}^{\dag
}d_m,  \nonumber \\
E_m =-\mu -2\left( l+1\right)t\cos\left(\frac{\pi f_m}{N}-\phi\right)
 +Ul.
\end{eqnarray}%
Setting $l$ and $\mu $ to zero, we realize the Hamiltonian derived in Sec. 3.

As an aside, this calculation actually yields more than just the description of
the quasi-angular momentum of a rotating ring lattice. In the limit of large $N$,
this system is indistinguishable from a linear lattice. Taking $t$ real, the 
Mott-insulator/superfluid phase diagram in two-state
approximation can be derived. Allowing the number of particles to
vary, it is apparent that the number of particles is determined by the
number of $m$'s for which $E_{m}<0$. Varying the chemical potential $\mu$, 
every time there exists an $m$ such that $E_{m}=0$, there is a phase
boundary between ground states of different numbers of particles. Thus, the
phase boundaries are given by the expression, 
\begin{equation}
\frac{\mu }{U}=-2\left( l+1\right) \frac{t}{U}\cos \left( \frac{\pi f_{m}}{N}\right) +l.
\end{equation}
In particular, it is possible to identify regions of integer filling and
non-integer filling, the boundaries between which are given by
\begin{equation}
\frac{\mu }{U}=\pm 2\left( l+1\right) \frac{t}{U}+l
\end{equation}
for a large number of sites. These boundaries yield an approximation to the
Mott-insulator superfluid phase diagram (figure A1). They have been derived 
in a slightly modified form in \cite{casetti2002vsc} using a semiclassical approach. We note that our
method can be easily extended to the case of superlattices, generating for
instance Mott phases of half-filling in the phase diagram. 
\begin{figure}[tbp]
\centerline{
\includegraphics[width=3in]{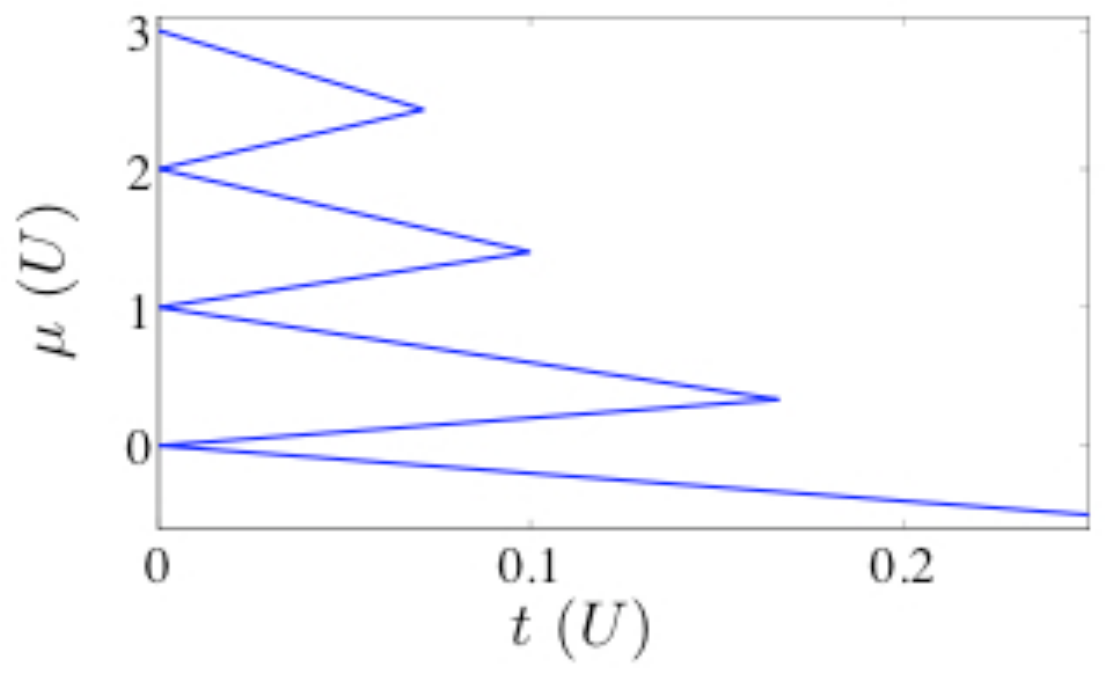}
}
\caption{(Color online.) Bose-Hubbard phase diagram for an infinite 
one-dimensional lattice in two-state approximation.}
\end{figure}

\section*{References}

\bibliographystyle{unsrt}
\bibliography{qambib}

\end{document}